\documentclass[pra,a4paper,twocolumn,superscriptaddress,longbibliography]{revtex4-2}
\usepackage{amssymb}
\usepackage{amsmath}
\usepackage{amsfonts}
\usepackage{graphicx}
\usepackage{bm}
\usepackage{xcolor}
\usepackage{multirow}
\usepackage{float}
\usepackage{comment}
\usepackage{ulem}
\usepackage{relsize}
\usepackage{cmap}
\usepackage{bbold}
\usepackage{physics}

\let\(\relax 

\let\)\relax

\newcommand{\eps}{\varepsilon}

\newcommand{\(}{\left(} 

\newcommand{\)}{\right)}

\DeclareMathOperator{\sgn}{sgn}

\DeclareMathOperator{\im}{Im}
\DeclareMathOperator{\re}{Re}

\begin{document}
	
	\title{Residual bulk viscosity of a disordered 2D electron gas}
	
	\author{V. A. Zakharov}
	
	\affiliation{Skolkovo Institute of Science and Technology, Moscow 121205, Russia}

	\author{I. S. Burmistrov}
	
	\affiliation{\hbox{L.~D.~Landau Institute for Theoretical Physics, acad. Semenova av. 1-a, 142432 Chernogolovka, Russia}}
	
	\affiliation{Laboratory for Condensed Matter Physics, HSE University, 101000 Moscow, Russia
	}

	\date{\today} 
	
	\begin{abstract}
		The nonzero bulk viscosity signals breaking of the scale invariance. We demonstrate that a disorder in a two-dimensional noninteracting electron gas in a perpendicular magnetic field results in the nonzero disorder--averaged bulk viscosity. We derive analytic expression for the bulk viscosity within the self-consistent Born approximation. This residual bulk viscosity provides the lower bound for the bulk viscosity of 2D interacting electrons at low enough temperatures.
	\end{abstract}

	\maketitle
	
	\section{Introduction}
	
	Hydrodynamic description of a viscous electron flow has a long history \cite{Gurzhi}. The progress in this field was detained by a lack of experiments (see, however, Ref. \cite{Molenkamp}). After experimental realization of graphene there were revival in the theoretical \cite{Muller2009,Andreev2011,Polini2015,Levitov2016,NGMS,Levitov2017,Kashuba2018,Lucas2018,Schmalian2019,Narozhny2019r,Narozhny2019r2} and experimental \cite{Titov2013,Polini2016,Kim2016,Moll2016,Geim2017,Levitov2018,Kvon2018,Berdyugin2019,Ilani2019,Ku2019,Vool2020} research on the hydrodynamic description of an electron transport in the two spatial dimensions.

	In the presence of the rotational symmetry the viscosity tensor of a two dimensional system can be parametrized by three parameters only,
	\begin{align}
		\eta_{jk,ls}  & = \bigl (\zeta- \eta_{s}\bigr )\delta_{jk}\delta_{ls} + \eta_{s} \bigl (\delta_{jl}\delta_{ks}+
		\delta_{js}\delta_{kl}\bigr ) \notag \\
		& + (\eta_H/2) \bigl (
		\epsilon_{jl}\delta_{ks}+\epsilon_{js}\delta_{kl}+\epsilon_{kl}\delta_{js}+\epsilon_{ks}\delta_{jl}\bigr ) .
		\label{eq:Visc:ten}
	\end{align}
	Here $\zeta$ stands for the bulk viscosity. The shear viscosity is denoted as $\eta_{s}$. The second line of Eq. \eqref{eq:Visc:ten} appears if the time reversal symmetry is broken, e.g. by a perpendicular magnetic field $B$. Similarly to the Hall conductivity, the Hall viscosity, $\eta_H$, describes the non--dissipative part of the viscosity tensor. We mention that the existence of the Hall viscosity has been well appreciated long time ago in the field of high temperature magnetized plasma~\cite{CC,Marshall,Kaufman,Thompson,Braginskii}.

	Although it is frequently said that viscosity exists only in a context of hydrodynamics, in fact, it  has  implication on its own: as a linear response that characterizes a change of the stress tensor under a time--dependent deformations~\cite{Bradlyn2012}. 
	In electron systems microscopic calculation of the viscosity tensor  has been traditionally performed for the shear and Hall components only~\cite{Steinberg,Avron,Levay,Avron1998,Read,RR,Haldane,HLF,Bradlyn2012,TH,Sherafati,Sherafati2019,Our-PRL,Liao2019,Liao2020,Rao2020}. The latter attracted much interest due to its relation to the geometrical response~\cite{Avron,Levay,Avron1998,Read,RR,Haldane,HLF} and quantization for translationally and rotationally invariant gapped quantum systems~\cite{RR}.  
	
	It is well-known that for a monoatomic  gas the Boltzmann kinetic equation predicts zero value for the bulk viscosity~\cite{Khalat,LP10}. Zero bulk viscosity implies that the system is scale invariant and can expand isotropically without dissipation. One more example of such a system is the unitary Fermi gas~\cite{Son2007}. However, generically, interaction breaks scale invariance and results in nonzero bulk viscosity. The canonical example is the Fermi liquid with nonzero albeit small bulk viscosity~\cite{Abrikosov1957,Sykes1970}. Recently, breaking of scale invariance has been extensively studied in the context of strongly interacting Fermi gas, both theoretically~\cite{Taylor2010,enss2011,Taylor2012,Hofmann2012,Dusling2013,Chafin2013,Enss2019} and experimentally~\cite{Holten2018,Peppler2018,Murthy2019}, as well as in the quantum chromodynamics~\cite{Karsch2008,Moore2008,Romatschke2009,Akamatsu2018}.

	Typically, a condensed matter electron system contains a quenched disorder. A presence of a random potential in the Hamiltonian inevitably breaks the scale invariance. Therefore, one may expect a nonzero value of the bulk viscosity even in the absence of electron-electron interactions. 
	
	In this paper, to unravel this issue, we consider a two-dimensional (2D) noninteracting electron gas in the presence of a perpendicular static magnetic field 
	and a random potential. Based on the Kubo formula for the bulk viscosity we demonstrate explicitly 
	how a nonzero magnitude of the disorder--averaged bulk viscosity appears due to the presence of a random potential in the Hamiltonian. We find that the real part of the bulk viscosity as a function of frequency contains two contributions: (i) a delta-function peak with the weight which is determined by such thermodynamic quantities as pressure 
	and isentropic compressibility; and (ii) a smooth part depending on the total elastic scattering time $\tau_0$. Within the self-consistent Born approximation (SCBA) we derive expression for the smooth contribution to the real part of the bulk viscosity at a finite frequency. In the absence of the magnetic field it acquires a remarkably simple form for all frequencies, $\omega$, and temperatures, $T$, much smaller than the chemical potential, $\mu$, 
	\begin{equation}
		\re \zeta(\omega) = \hbar^2 \nu_0/(2\tau_0) ,\qquad \hbar|\omega|, k_B T {\ll}\mu .
		\label{eq:BV:zeroFB}
	\end{equation}
Here $\nu_0$ denotes the density of states at the Fermi level. We emphasize that $\re \zeta(\omega)$ is proportional to the elastic scattering rate in contrast to the shear viscosity which, as other standard transport quantities, is proportional to the elastic scattering time. The result \eqref{eq:BV:zeroFB} indicates that in order to derive nonzero bulk viscosity within the kinetic equation approach one needs to take into account higher order corrections due to impurity scattering (see discussion in Sec. \ref{Sec:Summary}). The results reported in this paper together with the results of Ref. \cite{Our-PRL} provide the full answer for the viscosity tensor of 2D noninteracting electrons subjected to a perpendicular magnetic field.   
	
	The outline of the paper is as follows. In Sec.
	\ref{Sec:1} we formulate the problem and write down the Kubo-type expression for the bulk viscosity in which the delta-function peak at zero frequency is singled out. The weight of the delta-function peak is analyzed in Sec. \ref{Sec:delta}. In Sec. \ref{Sec:2} the SCBA is reviewed. We present calculation of the bulk viscosity within SCBA in Sec. \ref{Sec:3}. We end the paper with summary and conclusions (Sec. \ref{Sec:Summary}). Some details of calculations are given in Appendices.  Throughout the paper we use the units in which $\hbar=k_B=c=1$.

	\section{Formalism \label{Sec:1}}
	
	A 2D electron gas in the presence of an external static perpendicular magnetic field $B$ and a random potential $V(\bm{r})$ is described by the following Hamiltonian,
	\begin{equation}
		H = \bigl(-i\nabla - e \bm{A}\bigr )^2/(2m_e) + V(\bm{r}) .
		\label{eq:Hamiltonian}
	\end{equation}
	Here $m_e$ denotes the electron mass. The vector potential $\bm{A}(\bm{r})$ corresponds to the static magnetic field $B$, $\nabla\times \bm{A} = B \bm{e_z}$. We shall work in the Landau gauge: $A_y = B x$ and $A_x=A_z=0$. We assume the Gaussian distribution for a random potential with zero mean and characterized by the pair correlation function $\overline{V(\bm{r}) V(\bm{r^\prime})}= W(|\bm{r}-\bm{r^\prime}|)$. The function $W(r)$ is assumed to decay at a typical length scale $d_W$. The magnetic field $B$ is assumed to be strong enough to polarize the electron spins.

	In the microscopic theory the disorder-averaged viscosity tensor can be computed from the Kubo formula (see Eqs. (3.4), (3.11), and (3.14) of Ref. \cite{Bradlyn2012}):
	\begin{gather}
		\eta_{jk,ls}(\omega)\! =\! \frac{\delta_{jk}\delta_{ls}}{i\omega^+} (\kappa^{-1}\!-\!P) 
		\!-\!\!\! \int\!\! \frac{d \varepsilon f_\varepsilon}{\pi \mathcal{A} \omega^+}  \overline{\Tr [T_{jk},J_{ls}] \im G^R_\varepsilon}
		\notag \\
		+  \int \frac{d\varepsilon d\Omega}{\pi^2 \mathcal{A}} 
		\frac{\bigl (f_{\varepsilon}-f_{\varepsilon+\Omega}\bigr )}{i(\Omega-\omega^+)\omega^+} \overline{\Tr T_{jk} \im G^R_{\varepsilon+\Omega} T_{ls}\im
			G^R_\varepsilon} .
		\label{eq:Kubo:viscosity}
	\end{gather}
	Here $P$ stands for the \color{black} internal \color{black} pressure of the electron gas, $\kappa^{-1}$ denotes the inverse isentropic compressibility at constant particle number, $\mathcal{A}$ is the system area, and $\omega^+=\omega+i0$. 
	The retarded Green's function is defined in a standard way, $G^{R}_\varepsilon = 1/(\varepsilon-H+i0)$; $f_\varepsilon = 1/[1+\exp((\varepsilon-\mu)/T)]$ denotes the Fermi distribution function with the chemical potential $\mu$ and temperature $T$. The stress tensor operator $T_{jk}=m_e(v_jv_k+v_kv_j)/2$ is not affected by the presence of a random potential. Here $\bm{v} = (-i\nabla-e\bm{A})/m_e$ is the velocity operator \cite{Bradlyn2012,Our-PRL}. The strain generator operator $J_{jk}$ is related with the stress tensor operator as $T_{jk}=-i[H,J_{jk}]$. We note that contrary to the stress tensor operator, the expression for $J_{jk}$ is sensitive to the presence of a random potential. 
	\color{black} Disorder averaging in Eq.\eqref{eq:Kubo:viscosity} is denoted by an overbar. \color{black}

	Bulk viscosity $\zeta$ can be derived from the viscosity tensor by tracing the spatial indices, $\zeta =  \eta_{jj,ll}/d^2$, where $d=2$ is the spatial dimension. Using Eq.\eqref{eq:Kubo:viscosity}, we find
	\begin{gather}
		\zeta(\omega) = \frac{\kappa^{-1}-P-X}{i\omega^+} 
		+  \int \frac{d\varepsilon d\Omega}{(\pi d)^2 \mathcal{A}} 
		\frac{\bigl (f_{\varepsilon}-f_{\varepsilon+\Omega}\bigr )}{i(\Omega-\omega^+)\omega^+} 
		\notag \\
		\times \overline{\Tr T_{\Sigma} \im G^R_{\varepsilon+\Omega} T_{\Sigma}\im
			G^R_\varepsilon} ,
		\label{eq:Kubo:bulk-viscosity0}
	\end{gather}
	where $T_{\Sigma}=T_{jj}$ and the frequency independent quantity $X$ is defined as
	\begin{equation}
		X = i \int \frac{d \varepsilon f_\varepsilon}{\pi d^2 \mathcal{A}}  \overline{\Tr [T_{\Sigma},J_{\Sigma}] \im G^R_\varepsilon} .
		\label{eq:def:X}
	\end{equation}
	Here we introduce $J_{\Sigma}=J_{jj}$. Using the relation $T_\Sigma=2(H-V)$, we can rewrite Eq. \eqref{eq:Kubo:bulk-viscosity0} as follows
	\begin{gather}
		\zeta(\omega) = \frac{\kappa^{-1}-P-X}{i\omega^+} 
		+  4\int \frac{d\varepsilon d\Omega}{(\pi d)^2 \mathcal{A}} \frac{\bigl (f_{\varepsilon}-f_{\varepsilon+\Omega}\bigr )}{i(\Omega-\omega^+)\omega^+} 
		\notag \\
		\times \overline{\Tr V \im G^R_{\varepsilon+\Omega} V\im
			G^R_\varepsilon} .
		\label{eq:Kubo:bulk-viscosity}
	\end{gather}
	It is worthwhile to emphasize that the last term in the right hand side of the above expression represents the many-body two-point correlation function of a random potential. Thus the structure of Eq. \eqref{eq:Kubo:bulk-viscosity} resembles the structure of the Kubo formula for the interacting clean Fermi gas (see Ref. \cite{Nishida2020} and references therein). In our case a random potential plays a role of the contact operator \cite{Tan2008a,Tan2008b,Tan2008c}.
	
	The expression \eqref{eq:Kubo:bulk-viscosity} suggests the following sum rule for the disorder averaged bulk viscosity,
	\begin{equation}
		\int\limits_{-\infty}^\infty\frac{d\omega}{\pi} \zeta(\omega) = P+X-\kappa^{-1} .
		\label{eq:sum-rule}
	\end{equation}
	This expression is analogous to the sum rule found for the interacting clean Fermi gas \cite{Taylor2010,Nishida2020}.

	Using Eq. \eqref{eq:Kubo:bulk-viscosity}, we obtain the following Kubo formula for the real part of the bulk viscosity,  
	\begin{align}
		\re \zeta(\omega) = & \frac{4}{d^2}\int \frac{d\varepsilon}{\pi \mathcal{A}} 
		\frac{f_{\varepsilon}-f_{\varepsilon+\omega}}{\omega} 
		\overline{\Tr V \im G^R_{\varepsilon+\omega} V\im
			G^R_\varepsilon}
		\notag \\
		& +
		\pi \mathcal{D}\delta(\omega) ,
		\label{eq:Real-bulk-viscosity}
	\end{align}
	where the weight of the delta-function peak at $
	\omega=0$ is given as
	\begin{gather}
		\mathcal{D} = P+X-\kappa^{-1}-\color{black}\re \color{black}\frac{4}{d^2}\int \frac{d\varepsilon d\Omega}{\pi^2 \mathcal{A}} 
		\frac{f_{\varepsilon}-f_{\varepsilon+\Omega}}{\color{black}\Omega-i 0} 
		\notag \\
		\times
		\overline{\Tr V \im G^R_{\varepsilon+\Omega} V\im
			G^R_\varepsilon}.
		\label{eq:def:A:1}
	\end{gather}
	We emphasize that the appearance of a random potential $V$
	as vertices in Eq. \eqref{eq:Real-bulk-viscosity} reflects the fact that the bulk viscosity vanishes in the clean case.
	
	\section{The weight of the zero frequency delta-function peak \label{Sec:delta}}
	
\color{black} The expression for the weight \eqref{eq:def:A:1} involves the internal pressure which is proportional to the average value of the trace of the stress tensor, 
$P=\langle T_\Sigma\rangle/ (d \mathcal{A})$. We note that the presence of a random potential affects the standard relation for a Fermi gas between the internal pressure and the energy,
\color{black}
	\begin{gather}
		P = 
		\langle T_\Sigma\rangle/ (d \mathcal{A}) =
		- \int\frac{d\varepsilon}{\pi d \mathcal{A}} f_\varepsilon
		\overline{\Tr T_{\Sigma} \im G^R_\varepsilon} \notag \\
		= \frac{2}{d} \mathcal{E} +
		\frac{2}{d}\int\frac{d\varepsilon}{\pi \mathcal{A}} f_\varepsilon
		\overline{\Tr V \im G^R_\varepsilon}
		,
		\label{eq:Pressure-Energy}
	\end{gather}
	where we used the relation $T_\Sigma=2(H-V)$. 
	Here $\mathcal{E}=\int d\varepsilon \nu(\varepsilon) \varepsilon f_\varepsilon$ denotes the energy density where $\nu(\varepsilon)$ stands for the disorder-averaged density of states. We mention that the relation \eqref{eq:Pressure-Energy} is analogous to the Tan's relation for the pressure of an interacting Fermi gas \cite{Tan2008c}. In our case the random potential plays a role of the contact operator.
	
	\color{black} Next, using the relation 
	$[T_\Sigma,J_\Sigma] = 2 i T_\Sigma$ \footnote{This result can be derived from the expansion of the relation $e^{-2\lambda} (H-V) + V =e^{-i\lambda J_\Sigma} H e^{i\lambda J_\Sigma}$ to the second order in $\lambda$. See Ref. \cite{Bradlyn2012} for more details.}, we obtain
	\begin{equation}
		X = \frac{2}{d} P .
		\label{eq:expression:X:1}
	\end{equation}
Interestingly, this relation is not affected by the presence of a random potential. 	
\color{black}	
Using Eq. \eqref{eq:expression:X:1}, we rewrite the expression \eqref{eq:def:A:1} for the weight as
\color{black}
\begin{gather}
		\mathcal{D} = \frac{2+d}{d}P-\kappa^{-1}
-\frac{4}{d^2} \sum_{a\neq b} \frac{f_{E_a}-f_{E_b}}{E_a-E_b} |\langle a|V|b\rangle|^2 
.
		\label{eq:def:A:1:1}
	\end{gather}
Here $E_a$ and $|a\rangle$ denote the exact eigen energies and eigen states for the Hamiltonian $H$, $H|a\rangle=E_a| a\rangle$.
We note that the above expression for the weight $\mathcal{D}$ explicitly involves a random potential. \color{black} With the help of Eq. \eqref{eq:expression:X:1} the sum rule \eqref{eq:sum-rule} can be rewritten as 
\color{black}
	\begin{gather}
		\int\limits_{-\infty}^\infty\frac{d\omega}{\pi} \zeta(\omega) = \frac{2+d}{d}P-\kappa^{-1}
		 .
		\label{eq:sum-rule:2}
	\end{gather}
	\color{black}
	We note that the right hand side of Eq.\eqref{eq:sum-rule:2} is purely real \color{black}  and depends on the thermodynamic quantities only \color{black}.

	In the absence of the magnetic field and disorder, the inverse isentropic compressibility is defined as
	$\kappa^{-1} = - \mathcal{A}\bigl (\partial P/\partial \mathcal{A}\bigr )_{s \mathcal{A},n_e \mathcal{A}}$, 
	where $s$ and $n_e$ denote the entropy and electron densities, respectively. Using the thermodynamic relation $T s = \mathcal{E}+P-\mu n_e$ we find that a variation of the area $\delta \mathcal{A}$ under conditions $s \mathcal{A} = \textrm{const}$ and $n_e \mathcal{A} = \textrm{const}$ results in the following variation of the energy density, $\delta \mathcal{E} = - (\mathcal{E}+P)\delta \mathcal{A}/\mathcal{A}$. Also, a variation of the area leads to the variation of the electron density, $\delta n_e = - n_e \delta \mathcal{A}/\mathcal{A}$. Hence, we obtain \cite{Nishida2020}
	\begin{equation}
		\kappa^{-1} = (\mathcal{E}+P)\left (\frac{\partial P}{\partial \mathcal{E}}\right )_{n_e}    
		+n_e\left (\frac{\partial P}{\partial n_e}\right )_{\mathcal{E}} .
		\label{eq:InvKappa:1}
	\end{equation}
	We note that $\kappa^{-1}$ is related with the sound velocity, $c_s=1/\sqrt{\kappa m_e n_e}$. In the absence of the disorder, $V=0$, and the magnetic field, $B=0$, the energy density and the pressure of the ideal Fermi gas are related as  $P=2\mathcal{E}/d$ \cite{LP5}. This relation implies that the pressure is fixed if the energy density is fixed, i.e $(\partial P/\partial n_e)_{\mathcal{E}} \equiv 0$. Then, from Eq. \eqref{eq:InvKappa:1}, we find $\kappa^{-1}=(d+2) P/d$. As a result, we obtain that the weight of the delta-function peak is zero, $\mathcal{D}=0$. Therefore, Eq. \eqref{eq:Kubo:bulk-viscosity} implies that the bulk viscosity vanishes identically, $\zeta(\omega)=0$, for the ideal Fermi gas in agreement with its scale invariance.

	
	{\color{black} For clean 2D electron gas in presence of magnetic field  Eq. (\ref{eq:def:A:1:1}) simplifies to}
	\begin{equation}
\color{black}		\mathcal{D}=2P - \kappa^{-1}		 .
		\label{eq:D:mag:field}
	\end{equation}
	{\color{black} In that case internal pressure differs from ordinary thermodynamic pressure on the contribution associated with the action of the Lorentz force on the edge current and expressed as $P =-(\partial (\mathcal{EA})/\partial\mathcal{A})_{s\mathcal{A},n_e\mathcal{A},B}-\bm{m} \bm{B}\equiv -(\partial (\mathcal{EA})/\partial\mathcal{A})_{s\mathcal{A},n_e\mathcal{A},B\mathcal{A}}$, where $\bm{m}$ stands for the magnetization density \cite{LP8}. Isentropic compressibility}
	$\kappa^{-1} = - \mathcal{A}\bigl (\partial P/\partial \mathcal{A} \bigr )_{s \mathcal{A},n_e \mathcal{A},B \mathcal{A}}$
	{\color{black} is defined }at the constant particle number and the magnetic flux.
	
	Using the thermodynamic relation
	$T s = \mathcal{E}+P + \bm{m B}-\mu n_e$, we find that a variation of the area $\delta \mathcal{A}$ under conditions $s \mathcal{A} = \textrm{const}$, $n_e \mathcal{A} = \textrm{const}$, and $B \mathcal{A} = \textrm{const}$ results in the following variation of the energy density, $\delta \mathcal{E} = - (\mathcal{E}+P)\delta \mathcal{A}/\mathcal{A}$. Also, a variation of the area yields the variations of the electron density, $\delta n_e = - n_e \delta \mathcal{A}/\mathcal{A}$ and the magnetic field, $\delta B = - B \delta \mathcal{A}/\mathcal{A}$. 
	Hence, we obtain 
	\begin{gather}
		\kappa^{-1} = (\mathcal{E}+P)\left (\frac{\partial P}{\partial \mathcal{E}}\right )_{n_e,B}    
		+n_e\left (\frac{\partial P}{\partial n_e}\right )_{\mathcal{E},B}    
		\notag \\
		+ B\left (\frac{\partial P}{\partial B}\right )_{\mathcal{E},n_e} .
		\label{eq:InvKappa:2}
	\end{gather}
	Again, in the absence of a random potential, the weight of the delta-function peak vanishes. It is easy to check this statement at zero temperature. Then for $N$ filled Landau levels we find $P=\mathcal{E}=m\omega_c^2N^2/(4\pi)$ and $\kappa^{-1}=2\mathcal{E}$. Hence Eq. \eqref{eq:D:mag:field} leads to $\mathcal{D}=0$. 
	
	
	\section{Self-consistent Born approximation \label{Sec:2}}
	
	In order to take into account a random potential we employ  the self-consistent Born approximation~\cite{AFS}. This approximation is justified under the following conditions~\cite{RS,LA,DMP},
	\begin{equation}
		1/k_F, d_W {\ll} l_B,\qquad d_W{\ll} v_F \tau_0 .
		\label{eq:assumptions}
	\end{equation}
	Here $l_B=1/\sqrt{eB}$ stands for the magnetic length and $k_F=m_e v_F$ stands for the Fermi momentum with the Fermi velocity denoted as $v_F$. The total elastic relaxation time, $\tau_0$, in the absence of the magnetic field is defined by the following relation
	\begin{equation}
		\frac{1}{\tau_n}=\nu_0 \int\limits_{0}^{2\pi}\frac{d\phi}{2\pi} \tilde{W}\bigl (2k_F\sin(\phi/2)\bigr )\cos(n\phi), \, n=0,1,2, \dots    
	\end{equation}
Here $\tilde{W}(q)$ stands for the Fourier transform of $W(r)$. We note that the condition $k_F l_B{{\gg}}1$ is equivalent to the condition $N{{\gg}}1$ where $N$ is the number of filled Landau levels.
	
\begin{figure}[t]
		\centering\includegraphics[width=0.95\columnwidth]{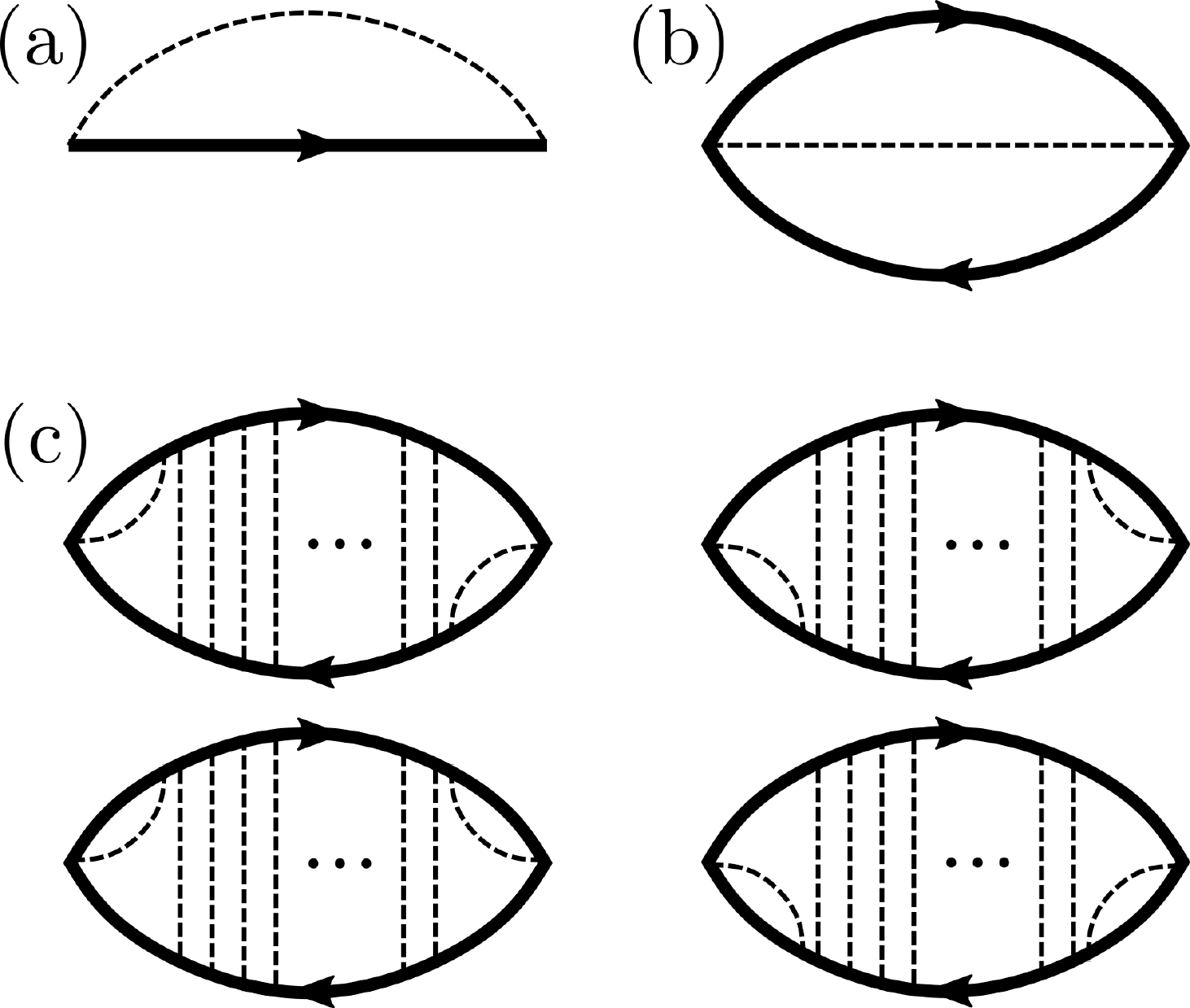} \caption{Diagrams used in SCBA. (a) The self-energy diagram;  (b) and (c) Diagrams corresponding to the bulk viscosity within SCBA. Bold solid lines denote the disorder averaged Green's function $\mathcal{G}_\varepsilon$, dashed lines stand for the pair correlation function $W(\bm{r})$.}
		\label{Figure:diagrams}
	\end{figure}

	Within the SCBA the physical quantities of interest are usually fully expressed in terms of the disorder averaged retarded Green's function $\mathcal{G}^R_\varepsilon$. It satisfies the self-consistency equation, see Fig. \ref{Figure:diagrams}(a),
	\begin{equation}
		\mathcal{G}^R_n=(\varepsilon-\epsilon_n-\Sigma^R_\varepsilon)^{-1}, \quad 
		\Sigma^R_\varepsilon = \frac{\omega_c}{2\pi \tau_0} 
		\sum_n \mathcal{G}^R_n ,
		\label{eq:self-consistent}
	\end{equation}
	where $\epsilon_n = \omega_c(n+1/2)$ denotes the energy of the $n$-th Landau level (LL) and $\Sigma^R_\varepsilon$ stands for the disorder averaged self energy. Here $\omega_c=eB/m_e$ is the cyclotron frequency. The self-consistency relation \eqref{eq:self-consistent} can be solved analytically for $\Sigma^R_\varepsilon$ in two limiting cases~\cite{AFS}. In the regime of a weak magnetic field, $\omega_c\tau_0{\ll} 1$, when LLs overlap, one can perform summation over LL index $n$ with the help of the Poisson formula and find \cite{AFS}
	\begin{equation}
		\Sigma^R_\varepsilon = -\frac{i}{2\tau_0} \left ( 1- 2\delta e^{2\pi i \varepsilon/\omega_c}\right ) ,
		\label{eq:SigmaR:weak}
	\end{equation}
	where 
	$\delta = \exp(-\pi/\omega_c\tau_0) {\ll} 1$ is the Dingle parameter. In the opposite case of well separated LLs, $\omega_c\tau_0{\gg} 1$, one can restrict the summation over LL index $n$ in Eq.~\eqref{eq:self-consistent} to $n=N$ only, where $\epsilon_N$ is the closest LL energy to the energy of interest: $|\varepsilon - \epsilon_N|<\omega_c/2$. Then one obtains~\cite{AFS}
	\begin{equation}
		\Sigma^R_\varepsilon = \frac{1}{2} \left (\varepsilon - \epsilon_N - i \sqrt{\Gamma^2 - 
			(\varepsilon - \epsilon_N)^2}\right ) .
		\label{eq:SigmaR:strong}
	\end{equation}
	Here the LL broadening is controlled by the energy scale $\Gamma = \sqrt{2\omega_c/(\pi \tau_0)}$. The disorder--averaged density of states can be expressed in terms of the disorder--averaged Green's function as
	\begin{equation}
		\nu_{\varepsilon} = -\frac{1}{2\pi^2 l_B^2} \sum_n \im \mathcal{G}^R_n(\varepsilon)=
		-2\tau_0\nu_0 \Im \Sigma^R_\varepsilon.
	\end{equation}
	Using Eqs. \eqref{eq:SigmaR:weak} and \eqref{eq:SigmaR:strong}, we find the disorder--averaged density of states \cite{AFS}
	\begin{equation}
		\nu_{\varepsilon} = \nu_0  \begin{cases}
			1- 2\delta \cos(2\pi \varepsilon/\omega_c), & \quad \omega_c\tau_0{\ll} 1, \\
			\tau_0 \sum\limits_n \re \sqrt{\Gamma^2-(\varepsilon-\epsilon_n)^2}, & \quad \omega_c\tau_0{\gg} 1 .
		\end{cases}
	\end{equation}

	\section{Bulk viscosity within SCBA\label{Sec:3}}
	
	The bulk viscosity at nonzero frequency, $\omega\neq 0$, is given by the first term in the right hand side of Eq. \eqref{eq:Real-bulk-viscosity}. We assume that frequency and temperature are much smaller than the chemical potential, $|\omega|,T{\ll} \mu$. Under this assumption, the integral over energy $\varepsilon$ is dominated by the vicinity of the chemical potential. 
	The unusual feature of the Kubo formula for the real part of the bulk viscosity, Eq. \eqref{eq:Real-bulk-viscosity}, is that vertex is a random potential. The diagrams contributing to $\re \zeta(\omega)$ within SCBA are shown in Fig.~\ref{Figure:diagrams}(b) and (c). 
	
	We start from computation of the diagram of Fig.~\ref{Figure:diagrams}(b). Using Eq. \eqref{eq:self-consistent}, we can rewrite this  contribution as
	\begin{align}
		\re \zeta^{(b)} & = \int \frac{d\varepsilon}{\pi} \frac{f_{\varepsilon}-f_{\varepsilon+\omega}}{\omega}\frac{1}{2\pi l_B^2}\sum\limits_{n}\Im\mathcal{G}^R_n(\varepsilon)\Im\Sigma^{R}_{\varepsilon+\omega}
		\notag\\
		& = \int d\varepsilon \frac{f_{\varepsilon}-f_{\varepsilon+\omega}}{\omega } \frac{\nu_{\varepsilon}\nu_{\varepsilon+\omega}}{2\tau_0\nu_0}.
		\label{eq:zeta:b}
	\end{align}
	We note that the contribution to $\re \zeta$ from the diagram of Fig.~\ref{Figure:diagrams}(b) can be expressed solely in terms of the density of states, $\nu_\varepsilon$, computed within SCBA.

	In addition to the diagram in Fig.~\ref{Figure:diagrams}(b) within SCBA one needs to take into account a set of diagrams shown in Fig.~\ref{Figure:diagrams}(c). They correspond to the impurity ladder insertion and describe vertex renormalization. As we shall see below, in spite of the scalar nature of the vertex (a random potential), the diagrams of Fig.~\ref{Figure:diagrams}(c) provide a significant contribution to the real part of the bulk viscosity in the case of a strong magnetic field. Evaluation of the four diagrams in Fig.~\ref{Figure:diagrams}(c) yields (see Appendix \ref{AppB})
	\begin{gather}
		\Re \zeta^{(c)} = \nu_0  \int d\varepsilon \frac{f_{\varepsilon}-f_{\varepsilon+\omega}}{\omega}
		\Re \Biggl[
		\frac{\left(\Sigma_{\varepsilon}^R+ \Sigma_{\varepsilon+\omega}^A\right)^2\Pi^{RA}_0(\omega)}{1-\Pi^{RA}_0(\omega)/\tau_0}
		\notag\\- 
		\frac{\left(\Sigma_{\varepsilon}^R+\Sigma_{\varepsilon+\omega}^R\right)^2 \Pi^{RR}_0(\omega)}{1-\Pi^{RR}_0(\omega)/\tau_0}
		\Biggr] .
		\label{eq:zeta:c}
	\end{gather}
	Here the polarization operator,
	\begin{equation}
		\Pi_0^{RA}(\omega)  = \frac{\omega_c}{2\pi}
		\sum_n \mathcal{G}^R_{n}(\varepsilon+\omega) 
		\mathcal{G}^{A}_{n}(\varepsilon) 
		=
		\frac{\tau_0(\Sigma^{A}_{\varepsilon}-\Sigma^R_{\varepsilon+\omega})}{\omega +\Sigma^{A}_{\varepsilon}- \Sigma^R_{\varepsilon+\omega}}
		,
		\label{eq:pol:oper}
	\end{equation}
	provides the contribution to the ``bubble'' without the impurity ladder insertion. The expression for $\Pi_0^{RR}(\omega)$ can be obtained from Eq. \eqref{eq:pol:oper} by changing superscript $A$ to $R$. Combining contributions \eqref{eq:zeta:b} and \eqref{eq:zeta:c}, we find the following expression for the disorder--averaged bulk viscosity of a 2D electron gas, 
	\begin{gather}
		\Re\zeta(\omega)= \int d\varepsilon \frac{f_{\varepsilon}-f_{\varepsilon+\omega}}{\omega} \frac{\nu_{\varepsilon}\nu_{\varepsilon+\omega}}{\tau_0\nu_0 }
		\Re
		\Biggl [
		\frac{1}{2} - \frac{ \Sigma^R_{\varepsilon+\omega}-\Sigma^{R}_{\varepsilon}}{\omega}
		\notag \\
		- \frac{\nu^2_{\varepsilon+\omega}-\nu^2_{\varepsilon}}{\omega \nu_{\varepsilon}\nu_{\varepsilon+\omega}}
		\bigl (
		\Sigma^R_{\varepsilon+\omega}+ \Sigma^{R}_{\varepsilon}
		\bigr )
		\Biggr ] .
		\label{eq:main:result}
	\end{gather}

	We emphasize that the above expression involves not only the density of states, $\nu_\varepsilon$, computed within SCBA but also the real part of the SCBA self energy. In the limit of zero frequency the expression \eqref{eq:main:result} becomes
	\begin{gather}
		\Re\zeta(\omega{\to}0)=\int \frac{d\varepsilon}{2 \tau_0\nu_0}  \bigl (- f^\prime_{\varepsilon}\bigr )\nu^2_{\varepsilon}
		\Bigl [
		1 - 2 \frac{\partial_{\varepsilon}
			\bigl (\nu^4_{\varepsilon} \Re \Sigma^R_{\varepsilon} \bigr )}{\nu^4_{\varepsilon}}
		\Bigr ] .
		\label{eq:main:result:00}
	\end{gather}
	In the absence of a magnetic field the density of states and self-energies are independent of energy. Therefore, Eq. \eqref{eq:main:result} transforms into remarkably  simple result \eqref{eq:BV:zeroFB}.
	It is instructive to compare the bulk viscosity and the shear viscosity in the absence of magnetic field \cite{Steinberg} and in the limit of zero frequency,
	\begin{equation}
		\frac{\Re\zeta(\omega{\to}0)}{\eta_s} = \frac{1}{\mu^2 \tau_{\rm tr,2}\tau_0} {\ll} 1
	\end{equation}
	where $1/\tau_{\rm tr,2} = 1/\tau_0-1/\tau_2$ denotes the inverse second transport time.
	
\begin{figure*}[t]
		\centerline{\includegraphics[width=0.8\textwidth]{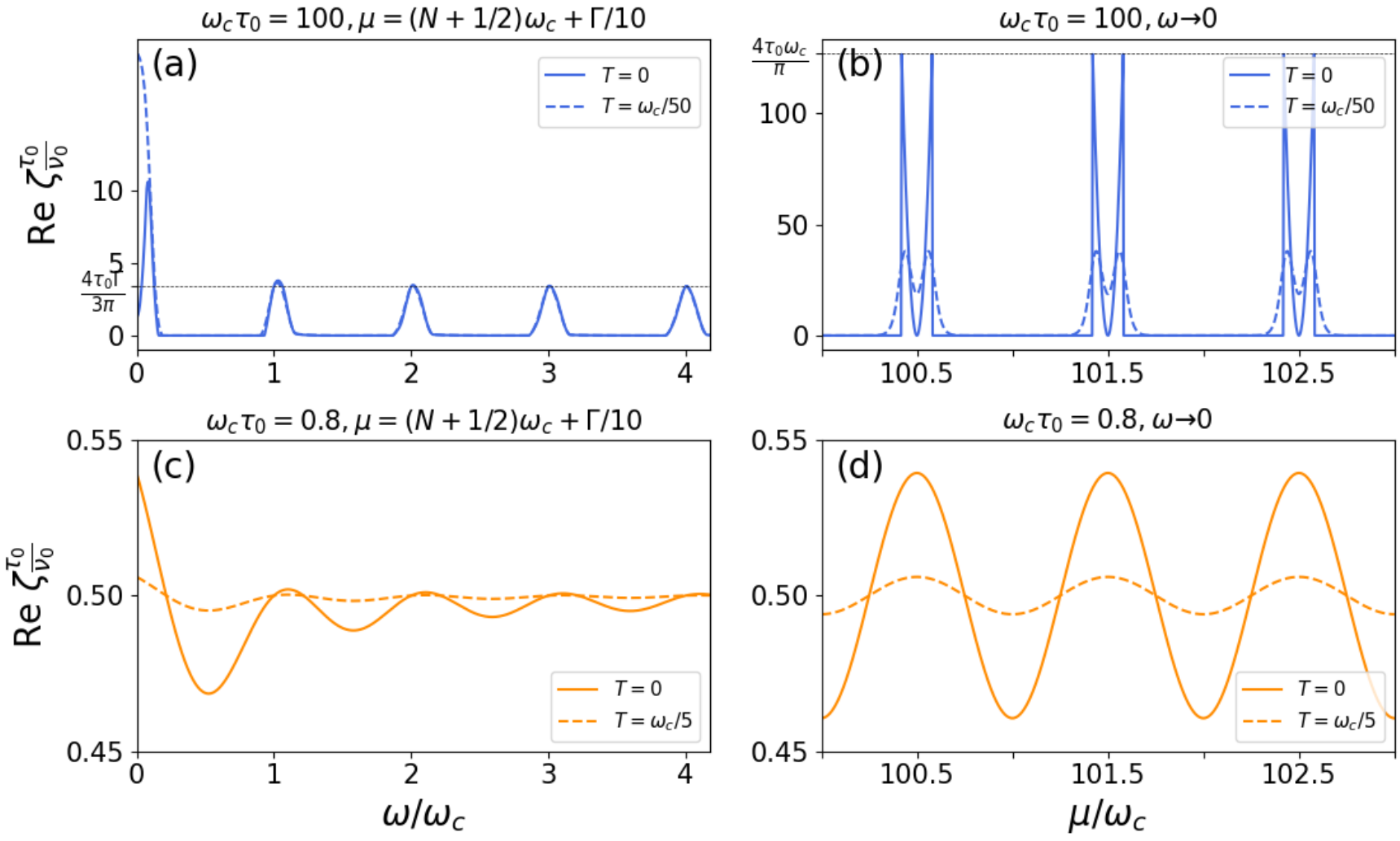}} \caption{The real part of the bulk viscosity in the regimes of strong ($\omega_c\tau_0=100$, panels (a) and (b)) and weak ($\omega_c\tau_0=0.8$, panels (c) and (d)) magnetic fields {\color{black} at zero (solid curves) and at finite (dashed curves) temperature}. The dependence of $\Re \zeta$ on frequency is shown on panels (a) and (c). Thin black dashed line at the panel (a) corresponds to the value of the bulk viscosity at the $\omega = k\omega_c$ for well-separated Landau levels, cf. Eq. \eqref{eq:BV:cyclotron}. Panels (b) and (d) show $\Re \zeta(\omega \to 0)$ as a function of the chemical potential for strong and weak magnetic fields. Thin black dashed line at the panel (b) corresponds to the limiting value of $\Re\zeta(\omega \to 0)$ at $T=0$ in the strong magnetic field for $\mu=\epsilon_N\pm \Gamma$, cf. Eq. \eqref{eq:BV:strong}.} 
		\label{Figure:zeta}
	\end{figure*}

	In the case of a weak magnetic field, $\omega_c\tau_0{{\ll}}1$, the general expression \eqref{eq:main:result} can be drastically simplified. To the first order in the Dingle parameter $\delta$ we find
	\begin{gather}
		\Re \zeta(\omega) = \frac{\nu_0}{2\tau_0} \Biggl[ 1 - 4 \delta \frac{\sin \Omega}{\omega}\left (1-\frac{2\pi}{\omega_c\tau_0} \frac{\tan(\Omega/2)}{\Omega}\right ) 
		\notag \\
		\times  \mathcal{F}_T
		\cos\frac{2\pi \mu}{\omega_c}
		\Biggr ] .
		\label{eq:BV:weak}
	\end{gather}
	Here $\mathcal{F}_T=(2\pi^2T/\omega_c)/\sinh(2\pi^2T/\omega_c)$ and $\Omega=2\pi \omega/\omega_c$ describe the temperature and frequency dependence of the Shubnikov--de Haas-type oscillations of the bulk viscosity, respectively. The real part of the bulk viscosity as a function of the frequency and the chemical potential in the case of a weak magnetic field is shown on   Fig.~\ref{Figure:zeta}(c) and (d). We mention that the amplitude of oscillations of $\Re\zeta(\omega)$ decays with the frequency  as $\sim \omega^{-1}$, while at zero frequency the amplitude of oscillations of $\Re\zeta$ with the chemical potential is independent of $\mu$. At zero temperature this amplitude is enhanced by the factor $\sim 1/(\omega_c\tau_0)$ in comparison with oscillations of the density of states. Therefore, the Shubnikov--de Haas-type oscillations in the bulk viscosity are stronger than in the longitudinal conductivity \cite{DMP,RevModPhys-ZudovPolyakovDmitriev} and the shear viscosity \cite{Our-PRL}. \color{black} Finite temperature suppresses the amplitude of oscillations of $\zeta$ with frequency and the chemical potential, see Fig. \ref{Figure:zeta}.\color{black}
	
	Now we consider the case of a strong magnetic field, $\omega_c\tau_0{{\gg}} 1$, in which Landau levels are well separated. Then in the limit of zero frequency the general result \eqref{eq:main:result:00} can be reduced to the following expression, 
\color{black}
	\begin{equation}
		\Re\zeta(\omega{\to}0)=\frac{2}{\pi^2l_B^2 \Gamma^2}\left [(\mu-\epsilon_N)^2+\frac{\pi^2T^2}{3}\right ], 
		\label{eq:BV:strong}
	\end{equation}
for $T, |\mu- \epsilon_N|{{\ll}} \Gamma$	and 
\begin{equation}
		\Re\zeta(\omega{\to}0)=\frac{\Gamma}{3\pi^2l_B^2 T}\left [1 - \frac{3\Gamma^2+5(\mu-\epsilon_N)^2}{20 T^2}\right ], 
		\label{eq:BV:strong:b}
	\end{equation}
for $|\mu- \epsilon_N|, \Gamma{{\ll}} T{{\ll}} \omega_c$. For large deviation of the chemical potential from the center of the Landau level, we find the zero-frequency bulk viscosity as
\begin{equation}
		\Re\zeta(\omega{\to}0) =
		\frac{2e^{-(\mu-\epsilon_N)/T}}{\pi^2l_B^2}
		\begin{cases}
		e^{\Gamma/T}, &  T{{\ll}} \mu-\epsilon_N-\Gamma ,\\
		\frac{2\Gamma}{3T} , &  \Gamma {{\ll}} T{{\ll}}  \mu-\epsilon_N .
		\end{cases}
		\label{eq:BV:strong:c}
	\end{equation}
	
	\color{black}

	The above results are a bit counterintuitive. At $T=0$ the real part of the bulk viscosity vanishes when the chemical potential is at the center of the $N$-th Landau level. With deviation of $\mu$ from the center of the Landau level $\Re\zeta(\omega\to 0)$ increases and reaches the magnitude  $2/(\pi l_B)^2$ at the boundary of the disorder broadened Landau level. This dependence on chemical potential is shown on  Fig.\ref{Figure:zeta}(b). Such unusual behavior of the real part of the bulk viscosity occurs since it is proportional to the derivative of the density of states with respect to the energy. \color{black}
	At nonzero $T$ a finite region of energies close to the chemical potential, $|\varepsilon - \mu| \lesssim T$, contributes to the integral over energies. Therefore the bulk viscosity increases with rising temperature if $\mu$ lies near the center of the Landau level and decreases if the chemical potential is situated near the band edge.
	\color{black}
		The maximum value of $\Re\zeta(\omega\to 0)$ is the factor $N^2 \tau_0/\tau_{\rm tr,2}$ smaller than the maximal value of the shear viscosity and the factor $N^2$ smaller than the maximal value of the Hall viscosity \cite{Our-PRL}.

	The result \eqref{eq:main:result} suggests that the bulk viscosity oscillates as a function of frequency with the period $\omega_c$. 
	Near harmonics of the cyclotron resonance, $|\omega-k\omega_c| = |\Delta \omega|{\ll}\omega_c$, $k=1,2,\dots$, Eq. \eqref{eq:main:result} transforms into the following expression
	\begin{equation}
		\Re\zeta(\omega) \approx  
		\int d\varepsilon \frac{f_{\varepsilon}-f_{\varepsilon+k\omega_c+\Delta\omega}}{k \omega_c} \frac{\nu_{\varepsilon}\nu_{\varepsilon+\Delta\omega}}{2\tau_0\nu_0 } .
	\end{equation}
	We note that at frequencies $\omega \gtrsim \omega_c$ one can neglect the terms with the self energy in the right hand side of Eq.~\eqref{eq:main:result}. 
	
	At zero temperature, $T=0$, and under assumption that $k$ is much smaller than the number of filled Landau levels, $N$, we obtain that the bulk viscosity near the $k$-th harmonics of the cyclotron resonance, $k=1,2,\dots $, is given by
	\begin{align}
		\Re\zeta(\omega) & = \frac{\Gamma \Theta\bigl (2\Gamma-|\Delta\omega|\bigr)}{2\pi^2l_B^2\omega_c}  
		\Biggl [
		\mathcal{F}_1\left(\frac{|\Delta\omega|}{\Gamma}\right )
		+ \frac{\sgn (\Delta\omega)}{k} 
		\notag \\
		& \times
		\Theta\bigl(\Gamma -|\mu-\epsilon_N|\bigr )
		\mathcal{F}_2\left(\frac{|\Delta\omega|}{\Gamma},\frac{\mu-\epsilon_N}{\Gamma}\right )
		\Biggr ].
	\end{align}
	Here $\Theta(x)$ stands for the Heaviside theta function and $\sgn (\Delta\omega)$ at $\Delta\omega=0$ is equal to zero. The functions $\mathcal{F}_{1,2}$ are defined as ($0\leqslant x < 2$, $0\leqslant |y| \leqslant 1$)
\begin{equation}
	\begin{split}
		\mathcal{F}_1(x) & = \int\limits_{-1}^{1-x} dt\, \sqrt{1-t^2}\sqrt{1-(t+x)^2} ,  \\
		\mathcal{F}_2(x,y) & = \int\limits_{\max\{y-x,-1\}}^{\min\{y,1-x\}} dt\, \sqrt{1-t^2}\sqrt{1-(t+x)^2} .
	\end{split}
	\end{equation}
	We mention that this result suggests that the magnitude of the bulk viscosity at the harmonics of the cyclotron resonance is independent of the harmonics number $k$ and the chemical potential, 
	\begin{equation}
		\Re\zeta(\omega=k\omega_c) =
		\frac{2 \Gamma}{3\pi^2l_B^2\omega_c} , \quad k=1,2,\dots .
		\label{eq:BV:cyclotron}
	\end{equation}
	As one can see, the magnitude of the bulk viscosity at the harmonics of the cyclotron resonance are the factor $\Gamma/\omega_c$ smaller than the maximal value of the bulk viscosity at small frequencies, $|\omega|{\ll} \Gamma$. Dependence of $\Re \zeta$ on frequency $\omega$ is shown on Fig. \ref{Figure:zeta}(a). 
	We note that the bulk viscosity decays relatively fast with detuning from the cyclotron resonance harmonics. 
	
	\color{black}
	The effect of nonzero temperature on the bulk viscosity at finite frequency can be described as follows. Temperature enters the factor $(f_\varepsilon - f_{\varepsilon+\omega})/{\omega}$ in the final expression for the bulk viscosity, see Eq. \eqref{eq:main:result}. Adjustments of this `weight' function are considerable only if $\varepsilon -\mu = \order{T}$ or $\varepsilon+\omega -\mu = \order{T}$. Hence, at large frequencies, $\omega {\gg} T$, the change of the `weight' function due to nonzero temperature is important for a small part of the energy integration region. Therefore, at $\omega {\gg} T$ the temperature does not significantly affect the bulk viscosity. At low but still nonzero frequencies the temperature effects are more significant. 
	\color{black}

	\section{Summary and conclusions \label{Sec:Summary}}
	
	To summarize, we have developed the theory of the disorder-averaged bulk viscosity of the disordered 2D electron gas in the presence of a perpendicular magnetic field within the self-consistent Born approximation. We demonstrated that the real part of the bulk viscosity has two contributions: delta-function peak at zero frequency, see Eq. \eqref{eq:D:mag:field}, and the smooth part, see Eq. \eqref{eq:main:result}. The latter is explicitly computed in the case of weak, see Eq. \eqref{eq:BV:weak}  and strong magnetic fields, see Eq. \eqref{eq:BV:strong}. Also we analyzed the harmonics of the cyclotron resonance in the case of strong magnetic fields, see Eq. \eqref{eq:BV:cyclotron}.

	\color{black}
	The zero field result \eqref{eq:BV:zeroFB} indicates that the method of the kinetic equation is not convenient for computation of the bulk viscosity. This statement is well enough illustrated by Ref. \cite{Sykes1970} where the bulk viscosity in the clean Fermi liquid was derived from the kinetic equation. One more example is calculations of the bulk viscosity of the clean interacting Fermi gas near the unitary limit within the kinetic equation approach \cite{Dusling2013,Chafin2013,Nishida2020}. However, it is worthwhile to explain for a reader how the kinetic equation can lead to the bulk viscosity which is proportional to the scattering rate, $1/\tau_0$ but not to the scattering time (as standard dissipative coefficients, e.g. the dissipative conductivity,  the shear viscosity, etc.). We start from expansion of the left hand side of the kinetic equation into formal series in $1/\tau_0$. Such an expansion can be symbolically written as $\mathcal{L}_0(n_q^{(0)}+\delta n_q)+\mathcal{L}_1 n_q^{(0)}+\dots$. Here $n_q^{(0)}$ denotes the equilibrium distribution function and  $\delta n_q$ stands for the out-of-equilibrium perturbation of the distribution function induced by a bulk flow of the electron gas. The operator $\mathcal{L}_{0}$ coincides with the operator in the kinetic equation for the clean noninteracting electron gas \cite{Khalat}. As a consequence, it vanishes acting on both  $n_q^{(0)}$ and $\delta n_q$. The operator $\mathcal{L}_{1}$ appears due to renormalization of the electron spectrum by scattering off a random potential, i.e., in other words, due to $\re \Sigma^R_\varepsilon$. Therefore, the term $\mathcal{L}_1 n_q^{(0)}$ is proportional to $1/\tau_0$. Since the collision integral is also proportional to $\delta n_q/\tau_0$, we find that the kinetic equation yields $\delta n_q \propto (1/\tau_0)^0$. This should be contrasted with a standard situation for which $\delta n_q \propto \mathcal{L}_{0}n_q^{(0)} (1/\tau_0)^{-1}$. Next, the bulk viscosity can be computed as $\zeta\propto \int d^2\bm{q}\, \mathcal{C}_q \delta n_q$ \cite{Sykes1970,Dusling2013,Chafin2013}. However, the function $\mathcal{C}_q$ becomes nonzero only due the renormalization of the electron spectrum by scattering off a random potential, i.e. $\mathcal{C}_q \propto 1/\tau_0$ (see similar cancellation for clean interacting problem \cite{Dusling2013,Chafin2013}). Again, we remind that in a standard case $\mathcal{C}_q$ is independent of $\tau_0$. Combining the estimates for $\delta n_q$ and $\mathcal{C}_q$, we find that the kinetic equation results in $\zeta\propto 1/\tau_0$.
	We emphasize that an actual computation of $\mathcal{L}_{1}$ and  $\mathcal{C}_q$, especially, in the presence of a magnetic field is much more complicated task than the diagrammatic approach developed in this work. 
	\color{black}
	
	We mention that the viscosity tensor affects the spectrum of bulk and edge magnetoplasmons \cite{Cohen2018}. Our result for the bulk viscosity in a weak magnetic field implies that the contribution to the magnetoplasmon spectrum due to the bulk viscosity can be neglected for wave vectors $q{\ll} k_F$ in comparison with the contribution due to the shear viscosity.  
	
	\color{black}
	It is instructive to estimate the magnitude of the bulk viscosity at zero magnetic field for a typical 2D electron gas in GaAs. In the absence of magnetic field the bulk viscosity at $T=0$ is given as  $\zeta = \nu_0/2\tau_0 =e/(4\pi \mu_u) \approx  10^{-18}$ g/s where we used the value of the mobility $\mu_u \approx 5\cdot 10^{4}$ cm$^{-2}$/(V $\cdot$ s). 
	For example, one may compare the above value of $\zeta$ with shear viscosity in the similar system $\eta =\hbar^2 \nu_0 \mu^2\tau / 2 = \hbar^2 n_e^2/(4\zeta)$ \cite{Our-PRL}, using the electron density $n_e \approx 10^{11}$ cm$^2$ one may found that $\eta \approx 10^{-15}$ g/s.
	Subsequently, this value may be compared with the value $10^{-12}$ g/s of the shear viscosity of electrons measured in graphene in the hydrodynamic regime \cite{Polini2016}. As one can see, in our regime the magnitude of the bulk viscosity is considerably smaller than the magnitude of the shear viscosity, which, in turn, is much smaller than typical shear viscosity in hydrodynamic regime. 
		\color{black}
	
	It is worthwhile to compare our result for the bulk viscosity due to a random potential with the result for the bulk viscosity in a clean weakly degenerate interacting Fermi gas. The interaction contribution to the bulk viscosity decreases with the temperature as a power law, \color{black} $\propto T^2$ \color{black} (see Ref. \cite{Sykes1970} for the three-dimensional Fermi liquid). This implies that the contribution to the bulk viscosity due to disorder dominates at low enough temperatures. Therefore, we expect that our results provide the lower bound for the residual bulk viscosity in 2D interacting disordered electron system at low temperatures. 
	
	The bulk viscosity can be estimated from measurements in interacting Fermi gases \cite{Holten2018,Peppler2018,Murthy2019}. It is an experimental challenge to extract the bulk viscosity from experiments in 2D electron systems. \color{black} There are two main difficulties for possible experimental measurement of the bulk viscosity of 2D electrons that we are aware of. The first issue is that a varying in time deformation should only be applied to the electronic system while impurities should not be affected. The second one is to measure experimentally the trace of the stress tensor. The first issue may be resolved as follows. One possibility is to use a quantum well in a semiconductor heterostructure with a $\delta$--layer (in which impurities are situated). Then one can apply a time-dependent deformation  only to semiconductor layers in which 2D electrons are formed. The other possibility is to use 2D electrons in van der Waals heterostructures with impurities situated in a substrate. Then again one can apply a slow time-dependent relative deformation $\delta\epsilon(t)=\epsilon \sin(2\pi f t)$ to a layer with 2D electrons only. To deal with the second problem we propose the following. To study the bulk viscosity one needs to measure the change in trace of the stress tensor due to an applied time-dependent deformation. Since the trace of the stress tensor is the internal pressure of the system, one can relate the change in the stress tensor with the change in the chemical potential at constant temperature, $\delta P = n_e \delta \mu$, which can be obtained from the Gibbs-–Duhem relation. Then the time-dependent  variation of the chemical potential, $\delta \mu(t)= (\pi f \zeta \epsilon/n_e) \cos(2\pi f t)$ can be measured, e.g. by technique similar to one reported in Ref. \cite{Yang2021}. 
Using the electron density $n_e = 10^{11}$ cm$^{-2}$,
frequency $f =1$ MHz, and deformation $\epsilon = 10^{-4}$, we obtain the amplitude of the change in the chemical potential of the order of 10$^{-11}$~K.   

Finally, we mention that our techniques can be extended to calculation of the bulk viscosity in a disordered graphene.	
	
	\color{black}

	\begin{acknowledgements}
		
		The authors are grateful to M. Goldstein and I. Gornyi for very useful discussions. The research was partially supported by the Russian Ministry of Science and Higher Education, the Russian Foundation for Basic Research (grant No. 20-52-12013) - Deutsche Forschungsgemeinschaft (grant No. EV 30/14-1) cooperation, and by the Basic Research Program of HSE. 
	\end{acknowledgements}
	
	\appendix

	\section{Ladder contribution to the bulk viscosity\label{AppB}}
	
	In this Appendix we present a brief derivation of Eq. \eqref{eq:zeta:c}. Diagrams corresponding to $\zeta^{(c)}$ are shown in Fig. \ref{Figure:diagrams}(c). We find
	\begin{gather}
		\Re \zeta^{(c)} = \frac{1}{\pi\mathcal{A}}\int d\varepsilon \frac{f_{\varepsilon}-f_{\varepsilon+\omega}}{\omega}
		\Biggl[
		X^{(1)}_{\varepsilon,\omega}+X^{(2)}_{\varepsilon,\omega}\notag\\+X^{(3)}_{\varepsilon,\omega}+X^{(4)}_{\varepsilon,\omega}
		\Biggr], 
	\end{gather}
	Here $X^{(1)}$ stands for the diagram in the upper left panel in Fig. \ref{Figure:diagrams}c, $X^{(2)}$ for the upper right panel, $X^{(3)}$ for the bottom left panel and $X^{(4)}$ for the bottom right panel, respectively. Each of the four diagram $X_i$ consists of three blocks: two self-energies at the vertices and the diffuson ladder in the middle. This ladder represents an infinite sum of diagrams with the Green's functions at the top and bottom, and arbitrary number of vertical dashed scattering lines. For computation of such diagrams it is convenient to rewrite $\overline{\Tr V\Im G^R_{\varepsilon+\omega} V \Im G^R_{\varepsilon}}$ as $-(1/4)\overline{\Tr V \left(G^R_{\varepsilon+\omega}-G^A_{\varepsilon+\omega}\right) V  \left(G^R_{\varepsilon}-G^A_{\varepsilon}\right) }$. After such transformation each contribution $X^{(i)}$ has four different terms with particular combination of the Green's functions. For the first contribution we obtain
	\begin{align}
		X^{(1)}_{\varepsilon,\omega} = -\frac{\pi \nu_0 \mathcal{A}}{2}\Sigma^{R}_\eps\Sigma^{R}_{\eps+\omega}\Pi^{RR}_0(\omega)\sum \limits_{n=0}^{\infty}\left(\frac{\Pi^{RR}_0(\omega)}{\tau_0}\right)^{n}\nonumber\\
		-\frac{\pi \nu_0\mathcal{A}}{2}\Sigma^{A}_\eps\Sigma^{A}_{\eps+\omega}\Pi^{AA}_0(\omega)\sum \limits_{n=0}^{\infty}\left(\frac{\Pi^{AA}_0(\omega)}{\tau_0}\right)^{n}\nonumber\\
		+\frac{\pi \nu_0\mathcal{A}}{2}\Sigma^{R}_\eps\Sigma^{A}_{\eps+\omega}\Pi^{RA}_0(\omega)\sum \limits_{n=0}^{\infty}\left(\frac{\Pi^{RA}_0(\omega)}{\tau_0}\right)^{n}\nonumber\\
		+\frac{\pi \nu_0\mathcal{A}}{2}\Sigma^{A}_\eps\Sigma^{R}_{\eps+\omega}\Pi^{AR}_0(\omega)\sum \limits_{n=0}^{\infty}\left(\frac{\Pi^{AR}_0(\omega)}{\tau_0}\right)^{n}\notag\\
		= -\frac{\pi \nu_0\mathcal{A}}{2}\Biggr[\frac{\Sigma^{R}_\eps\Sigma^{R}_{\eps+\omega}\Pi^{RR}_0(\omega)}{1-\Pi^{RR}_0(\omega)/\tau_0}+ \frac{\Sigma^{A}_\eps\Sigma^{A}_{\eps+\omega}\Pi^{AA}_0(\omega)}{1-\Pi^{AA}_0(\omega)/\tau_0} \nonumber\\
		-\frac{\Sigma^{R}_\eps\Sigma^{A}_{\eps+\omega}\Pi^{RA}_0(\omega)}{1-\Pi^{RA}_0(\omega)/\tau_0}- \frac{\Sigma^{A}_\eps\Sigma^{R}_{\eps+\omega}\Pi^{AR}_0(\omega)}{1-\Pi^{AR}_0(\omega)/\tau_0}\Biggl] .
	\end{align}
	
	As one can check,  $X^{(2)}_{\eps,\omega}=X^{(1)}_{\eps,\omega}$. Next, in a similar way, we find
	\begin{align}
		X^{(3)}_{\eps,\omega} =  -\frac{\pi \nu_0 \mathcal{A}}{2}\Biggr[\frac{\bigl(\Sigma^{R}_\eps\bigr )^2\Pi^{RR}_0(\omega)}{1-\Pi^{RR}_0(\omega)/\tau_0}+ \frac{\bigl(\Sigma^{A}_\eps\bigr)^2\Pi^{AA}_0(\omega)}{1-\Pi^{AA}_0(\omega)/\tau_0} \nonumber\\
		-\frac{\bigl (\Sigma^{R}_\eps\bigr )^2\Pi^{RA}_0(\omega)}{1-\Pi^{RA}_0(\omega)/\tau_0}- \frac{\bigl(\Sigma^{A}_\eps\bigr )^2\Pi^{AR}_0(\omega)}{1-\Pi^{AR}_0(\omega)/\tau_0}\Biggl]
	\end{align}
	and
	\begin{align}
		X^{(4)}_{\eps,\omega} =  -\frac{\pi \nu_0 \mathcal{A}}{2}\Biggr[\frac{\bigl(\Sigma^{R}_{\eps+\omega}\bigr)^2\Pi^{RR}_0(\omega)}{1-\Pi^{RR}_0(\omega)/\tau_0}+ \frac{\bigl(\Sigma^{A}_{\eps+\omega}\bigr)^2\Pi^{AA}_0(\omega)}{1-\Pi^{AA}_0(\omega)/\tau_0} \nonumber\\
		-\frac{\bigl (\Sigma^{A}_{\eps+\omega}\bigr )^2\Pi^{RA}_0(\omega)}{1-\Pi^{RA}_0(\omega)/\tau_0}- \frac{\bigl (\Sigma^{R}_{\eps+\omega}\bigr )^2\Pi^{AR}_0(\omega)}{1-\Pi^{AR}_0(\omega)/\tau_0}\Biggl] .
	\end{align}
	Combining these four contributions together, one can derive Eq. \eqref{eq:zeta:c}.

	
%

\end{document}